# Calculation of thermal expansion coefficient of Fe$_3$Al with the addition of transition metal elements

Tatiana Seletskaia, Leonid Muratov, Bernard Cooper,
West Virginia University, Physics Departement,
Hodges Hall, PO Box 6315, Morgantown, WV 26506-6315

**Abstract**

The addition of transition metal elements can significantly modify physical properties of intermetalic compounds. We studied the influence of Molybdenum and Vanadium additives on thermal expansion coefficient (CTE) of Fe$_3$Al and FeAl over the wide range of temperatures. The site preference of both transition metals was determined by full-potential LMTO method within the grandcanonical formalism. At low temperatures CTEs were found directly from the FP-LMTO calculations by incorporating them into the Debye model of a solid. The obtained thermal expansion for pure Fe$_3$Al and FeAl is within 10% of its experimentally measured values. At high temperatures we performed molecular dynamics simulations based on our many-body atomistic potentials. The parameters were fitted to reproduce the total energy of a crystal under various types of deformations obtained by FP-LMTO method and were tested with respect to different structures and vacancy formation energies. Our calculations show that addition of V decreases the CTEs of both iron-aluminides, while the addition of Mo makes Fe$_3$Al DO$_3$ structure unstable.

**Introduction**

Recent experiments have shown that addition of Mo and V leads to a remarkable stabilization of Fe$_3$Al DO$_3$ structure and therefore improves its strength [1]. Thus, there is a natural interest in studying the effect of these additives on the other physical properties of iron-aluminides. Fe$_3$Al in DO$_3$ structure has three non-equivalent lattice sites: Fe$_I$ site surrounded by eight iron atoms, Fe$_{II}$ site with four iron and four aluminum nearest neighbors and Al site. Since experimental data on site-selection of Mo and V in Fe$_3$Al is still incomplete [1], we solved this problem using FP-LMTO method.

One of the most important characteristics of iron-aluminides is the strong covalent bonding between Fe and Al atoms implying high activation energy of the optical phonons. Thus, thermal expansion of these compounds can be calculated directly by FP-LMTO method incorporated into the Debye model of a solid. This model was succesfully applied by C.L.Fu *et al* to calculate anisotropic CTEs of molybdenum-silicides [2]. We tested the validity of the Debye model for pure iron-aluminides and in the presence of Mo and V additives by comparing the calculated thermal expansion with the one obtained from MD simulations. Further, we studied the effect of Mo and V addition on thermal expansion of FeAl and Fe$_3$Al.

**Methodology**

*FP-LMTO method*

A complete description of FP-LMTO method can be found in [3]. It is a highly accurate technique for solving bulk density functional problem within Local Density Approximation (LDA). Our FP-LMTO method has a full potential both in the muffin-tin and interstitial regions. In addition, the muffin-tin orbitals are not constrained to have zero kinetic energy in the interstitial area. These two features of FP-LMTO method are essential for calculating the total-energy and electronic structure in low-symmetry crystal systems. Multiple kappas used in all calculations, provide basis enrichment that allows capturing the behavior of *3d* and *4d* electrons.

*Many-body atomistic potentials*

To test the applicability of the Debye model to iron-aluminides, we were using molecular dynamics simulation with many-body atomistic potentials [4]. We divide the total energy of a crystal into the following three parts: many-body interaction energy, pair-potential and three-body interaction energy:

$$E = \sum_i e_{si}(\rho_i, \{p_1\}) + \sum_{i \neq j} V_2(r_{ij}, \{p_2\}) + \sum_{i, j \neq i, k \neq i} V_3(r_{ij}, r_{ik}, \{p_3\}), \quad (1)$$

where $\rho_i$ is the atomic density, $r_{ij}$ are the distances between the atoms and $\{p_1\}$, $\{p_2\}$ and $\{p_3\}$ are the sets of parameters obtained by fitting the results of FP-LMTO calculations.

The atomic density of a pure metal with a given volume $\omega$ is defined as a relationship of this volume to the volume of the equilibrium $\omega_e$ and is written as follows:

$$\rho_i = \frac{\omega_e}{\omega} = f_i + \sum_j \xi_j(r_{ij}) \quad (2)$$

Here, on the right-hand side $f_i$ is the atomic type dependent parameter, the summation is carried over all neighbor atoms of the *i-th* atom at a distance given by $r_{ij}$ and $\xi_j(r_{ij})$ is some parameterized function. The following form of $\xi_j(r_{ij})$ is found to be the simplest and the most flexible for fitting atomic densities of *bcc, fcc* and *hcp* structures simultaneously:

$$\xi_j(r_{ij}) = q r_{ij}^k [\exp(-\zeta r_{ij}) - y \exp(-\eta r_{ij})] \quad (3)$$

There are six atomic type dependent parameters including $f_i$, which are determined by uniformly expanding and contracting the crystal lattice of a selected reference structure.

The functional form of the many-body interaction energy in (1) should be rich enough to describe all possible deformations of the system and therefore, we define this functional implicitly through its relationship to the uniform expansion energy curve $E(\omega)$:

$$N \cdot e_{si}(\rho_i, \{p_1\}) = E(\omega) - \sum_{i \neq j} V_2(r_{ij}, \{p_2\}) - \sum_{i, j \neq i, k \neq i} V_3(r_{ij}, r_{ik}, \{p_3\}), \quad (4)$$

where $N$ is the total number of the atoms. Since the crystal symmetry is completely preserved during uniform expansion, the interatomic distances $r_{ij}$ can be written as the functions of the atomic density. We parameterized the uniform expansion curve $E(\omega)$ in the following form:

$$E(\omega)/N = e_0 - 2 e_1 \exp(-\alpha \frac{\omega_e}{\omega}) + e_2 \exp(-2\alpha \frac{\omega_e}{\omega}) \quad (5)$$

The pair-potentials in (1) and (4) are the functions of the interatomic distance:

$$V_2(r_{ij}) = \phi_0 [1 + s_1(r_{ij}/r_0 - 1)] \exp[-s_2(r_{ij}/r_0 - 1)] \quad (6)$$

The angular dependence of the atomic bonding is described by three-body interaction which written as follows:

$$V_3 = \frac{1}{2} \sum_{j \neq i, k \neq i} w(r_{ij}) w(r_{ik}) (\cos^2 \Theta_{ijk} - \frac{1}{3}), \quad w(r_{ij}) = g_{ij} \exp(-\frac{\gamma_{ij}}{r_c - r_{ij}}), \quad (7)$$

where $r_{ij}$, $r_{ik}$, are the radius vectors of the atoms *j* and *k* drawn from the *ith* atom as a center, $\Theta_{ijk}$ is the angle between these radius vectors.

The interaction cutoff function and the parameters for FeAl and Fe$_3$Al with Mo and V additives can be found in the Appendix.

**Results**

*a) Site-preference of Mo and V additives in Fe$_3$Al*

To calculate site-selection energy of Mo and V in Fe$_3$Al we considered 32-atom supercell. One of the atoms at the origin was substituted by the impurity. The atomic positions after lattice relaxation were found from the forces calculated by Hellmann-Feynman theorem. In our calculations we used a common definition of the defect formation energy:

$$E^f = TE_{defect} + \mu_{defect} - TE_{perfect} , \qquad (8)$$

where $TE_{defect}$ is the total energy of a supercell with a defect; $TE_{perfect}$ is the total energy of a corresponding perfect supecell; $\mu_{defect}$ is the chemical potential defined as the difference between the total energies of original and substituted atoms being in a bulk.

The results of calculations are presented in Table 1. The relaxation energy and change in the bond-length are defined with respect to the perfect Fe$_3$Al 32-atom supercell. Site-selection energy is defined with respect to the smallest formation energy of a given defect. For both Mo and V it is transition metal substituted at Fe$_I$ site.

As shown in Table 1, the relaxation energies of Mo in Fe$_3$Al are significantly larger than their V counterparts due to the large atomic radius of Mo. By examining the site-selection energies, we can conclude that the probability of Mo to be found at Fe$_I$ site is slightly larger than for the other sites.

The addition of V to Fe$_3$Al leads to a moderate relaxation of the crystal lattice or no relaxation at all if placed at Fe$_I$ site. The total energy of the supercell with V atom located at Fe$_I$ site after relaxation is 2018K/atom smaller than the supercell energy with V placed at Fe$_{II}$ site. This result is in a good agreement with the calculations of Reddy *et al* [5] although their calculations were performed for a different model and using another computational technique. Their obtained site-selection energy between Fe$_I$ and Fe$_{II}$ sites is 1870K/atom.

Also, V additives are usually used to compensate for the decrease in the iron content, we considered the possibility of Al atoms to be substituted by the transition metal additives because Al->Fe$_I$ antisite defect energy is close to zero [6]. Thus, we found that the site-selection energy between Fe$_I$ and Al sites is almost half of one obtained for the two different iron sites.

**Table 1** Relaxation and site-selection energies of Mo and V defects in Fe$_3$Al.

| System | Relaxation energy, eV/cell | Change in the bond length | Site-selection energy, K/atom |
|---|---|---|---|
| Mo-Fe$_I$ | 0.10 | 0.5% | 0.00 |
| Mo - Fe$_{II}$ | 0.41 | 2.9% | 25 |
| Mo – Al | 0.077 | 0.1% | 741 |
| V- Fe$_I$ | No relaxation | 0.0% | 0.00 |
| V - Fe$_{II}$ | 0.02 | 0.4% | 2018 |
| V - Al | 0.025 | -0.8% | 1231 |

The experimental results on the site-selection of V in $Fe_3Al$ are quite contradictory. Nishino *et al* used X-ray diffraction technique to study the dependence of lattice parameter on V concentration $x$, in $(Fe_{1-x}V_x)_3Al$ compound [1]. They found that the lattice constant decreases with the growing $x$ and reaches its minimum at about *33.3%* of concentration. This result was regarded as an experimental evidence for the $Fe_I$ site selection of V at least for this composition range. Another research group reported the results of X-ray diffraction measurements for $Fe_2VAl$ crystals [7]. They observed appreciable antisite disorder in all of their samples. Our computational results speak in favor of the first X-ray diffraction measurements, since the obtained site-preference energy between $Fe_I$ and $Fe_{II}$ sites is relatively large. However, the site-selection energy between $Fe_I$ and Al sites is comparable to temperatures involved in the metallurgical synthesis of this iron-aluminide. Thus, we can conclude that the actual site occupation of V may depend significantly on the details of alloy formation.

### *b) Calculation of thermal expansion of pure iron-aluminides*

Linear thermal expansion of pure FeAl and $Fe_3Al$ was calculated by two different methods: using the Debye model of a solid and by MD simulation. The obtained results for FeAl and $Fe_3Al$ are shown in Figures 1 (a) and (b), respectively. For both iron-aluminides we got a good agreement between the Debye model and MD indicating the validity of the former approximation for these compounds. The difference becomes noticeable only at high temperatures due to the growing contribution of the optical phonons to the thermal expansion and anharmonic effects being excluded in the Debye model. For $Fe_3Al$ we obtained a very good agreement between theory and experiment up to the temperature 600K. At the temperature 830K a second-order phase transition of $Fe_3Al$ from $DO_3$ structure to disordered $B_2$ structure takes place. Since our model does not account for this, there is an increasing divergence of theoretical and experimental results.

**Figure 1.** Linear Thermal Expansion of FeAl and $Fe_3Al$. The solid line with circles represents the results of MD simulation, the dashed line shows the thermal expansion obtained within the Debye model and we used triangles to denote the experimentally measured thermal expansion from [8].

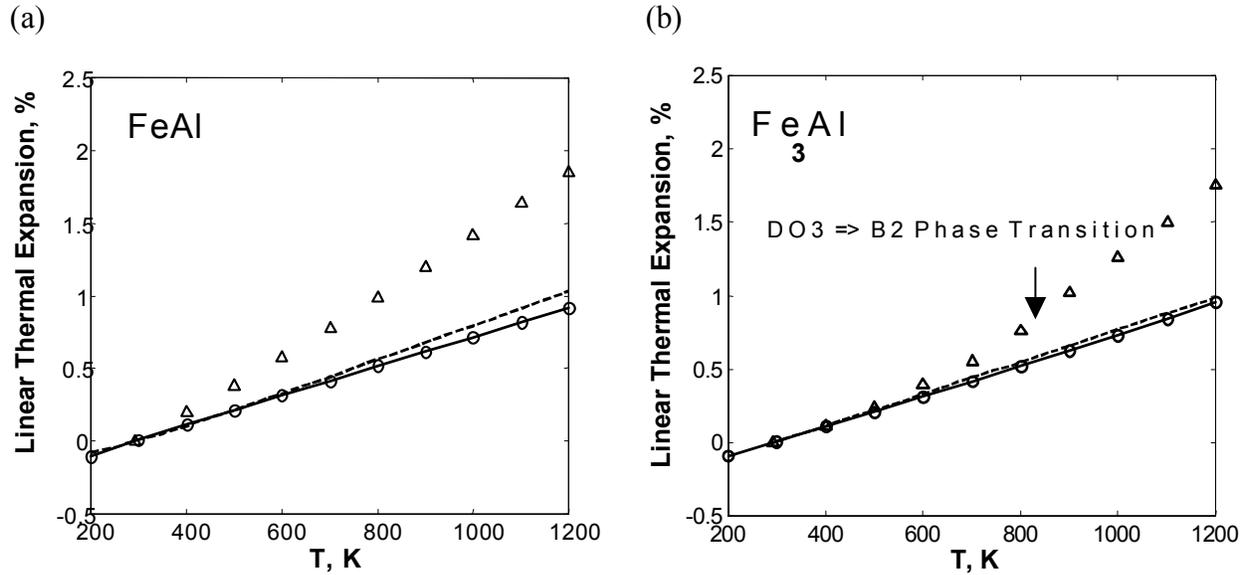

*c) Thermal expansion of iron-aluminides in the presence of transition metal additives*

We calculated linear thermal expansion of $(Fe_{0.917}V_{0.083})_3Al$ compound using both the Debye model of a solid and by MD simulation. To avoid additional errors, the interaction between the atoms in the both cases was described by many-body atomistic potentials. The comparison between two calculations is presented in Figure 2. One can see that the Debye model significantly overestimates the thermal expansion in the case of the additives. The existence of impurities in crystal destroys the lattice symmetry and changes the phonon spectrum by lowering the optical modes. Thus, their excitation requires less energy and the atomic oscillations become less organized resulting in a smaller thermal expansion than the one obtained within the Debye model.

The dependence of CTEs of $Fe_{1-x}V_xAl$ and $(Fe_{1-x}V_x)_3Al$ pseudo-binary alloys on V concentration, $x$, is shown in Figures 3 (a) and (b), respectively. One can see that in both cases CTE decreases with the increase of $x$. While CTE of FeAl preserves its extremely small temperature dependence with the addition of V, CTE of $(Fe_{1-x}V_x)_3Al$ alloy changes from merely temperature-independent at 4.3% and 6.5% of V to being dependent on the temperature at $x=8.3\%$.

**Figure 2** Linear Thermal Expansion of $(Fe_{1-x}V_x)_3Al$ pseudo-binary alloy (*x=8.3%*).

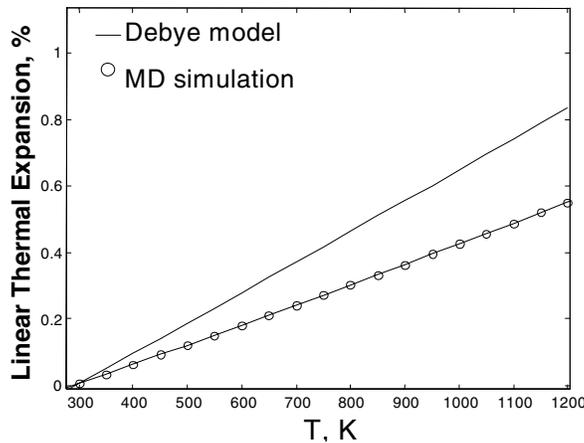

**Figure 3** CTE of $Fe_{1-x}V_xAl$ and $(Fe_{1-x}V_x)_3Al$ alloys. The results of MD simulation.
(a) (b)

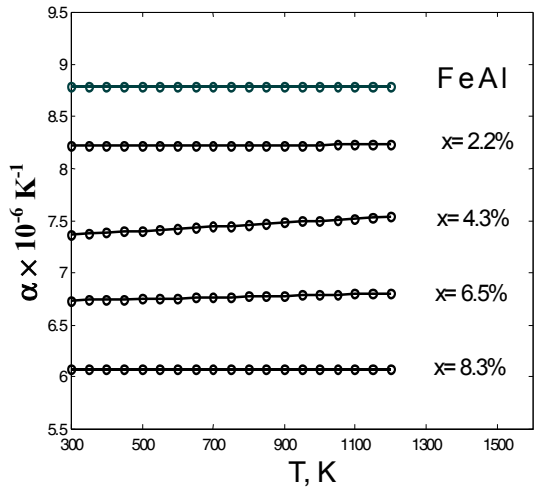 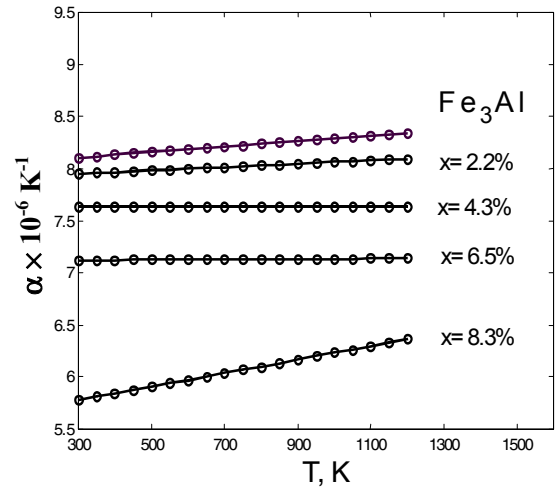

**Figure 4** CTE of $(Fe_{1-x}Mo_x)_3Al$ pseudo-binary alloys. The results of MD simulation.

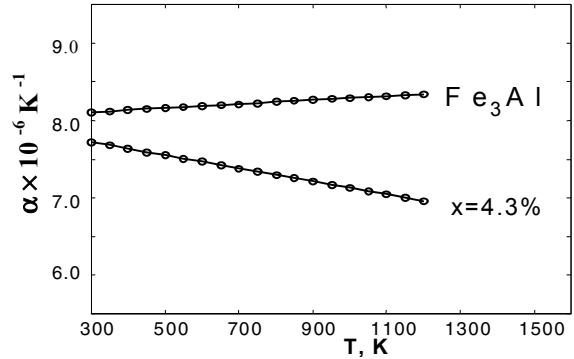

To study the effect of Mo additives, we considered only one concentration of this transition metal in $Fe_3Al$ compound, namely 4.3% of Mo substituted for the Fe content. The calculated CTE, presented in Figure 4, decreases with temperature indicating the structural instability of the compound. A similar effect of Mo addition was observed in nickel binary alloys [9].

**Conclusions**

We have considered site-selection problem for Mo and V in $Fe_3Al$ using FP-LMTO method. Our calculations show that $Fe_{II}$ site is energetically more favorable for both transition metals. The obtained site-selection energies of Mo are relatively small, while for V they are much larger indicating that this additive is likely to improve the stability of $Fe_3Al$ $DO_3$ structure.

We calculated thermal expansion of pure iron-aluminides and with the addition of V and Mo. For pure iron-aluminides we obtained a reasonably good agreement with the experiment. The thermal expansion calculated within the Debye model differs from the results of MD only at high temperatures. However, when the additives are present, the Debye model does not provide the correct description of thermal expansion. Our MD calculations showed that addition of V decreases CTE of both FeAl and $Fe_3Al$, while the addition of Mo makes $DO_3$ of $Fe_3Al$ unstable.

# Appendix

*a) Cutoff function.* In practice, the values of potential parameters depend on the cutoff distance of the interaction. To minimize this effect, the cutoff distance has to be situated in between the neighboring lattice shells. Unfortunately, the lattice constants of FeAl and Fe$_3$Al are very different from the lattice constants of pure Mo and V metals. Therefore, we have chosen cutoff distance to be dependent on the type of the atoms. The cutoff function for many-body interaction and pair-potential has the following distance dependence:

$$F_{cutoff}(x) = (1-x)^3(1+3x+6x^2), x = \frac{r_{ij}-r_b}{r_c-r_b}, \qquad (9)$$

where $r_c$ is the cutoff distance and $r_b$ is the point, where cutoff function gets connected with the density function. Such a form of the cutoff function guarantees that the first and the second-order derivatives of $F_{cutof}(r_{ij})$ are equal to zero at $r_b$ and $r_c$. For the atoms of the same type the cutoff distances are listed in Table 2. The interaction between the atoms of different types takes place only in the compounds. The cutoff distance for the interaction between different atoms was chosen to be in between fifth and sixth neighboring shells ($r_b$ =5.10A, $r_c$ =5.20A) that resulted in taking into account up to 58 next-neighbors in B$_2$ and DO$_3$ structures.

*b) Many-body atomistic parameters for iron-aluminides in the presence of transition metal additives.*

**Table 2** Cutoff distances in A for atomic density function (2) and pair-potentials (6) for the atoms of the same type.

|       | Al   | Fe   | V    | Mo   |
|-------|------|------|------|------|
| $r_b$ | 5.10 | 5.10 | 5.30 | 5.70 |
| $r_c$ | 5.20 | 5.20 | 5.40 | 5.80 |

**Table 3** Atomic density parameters found from the fitting of the relationship given by (3). The interatomic distances are scaled by $r_e$, where $r_e$ is the bond length of the pure metals. The parameters are dimensionless.

| Type of Atom | $r_e$, A | $f$    | $q$   | $\zeta$ | $k$   | $y$   | $\eta$ |
|--------------|----------|--------|-------|---------|-------|-------|--------|
| Al           | 2.75     | 0.4034 | 55.85 | 2.826   | 0.303 | 0.967 | 2.828  |
| Fe           | 2.337    | 0.2264 | 53.76 | 2.844   | 0.953 | 0.969 | 2.830  |
| V            | 2.52     | 0.2358 | 54.13 | 2.844   | 0.908 | 0.969 | 2.830  |
| Mo           | 2.704    | 0.2443 | 54.33 | 2.844   | 0.853 | 0.969 | 2.830  |

**Table 4** Potential parameters for lattice expansion energy curve $E(\omega)$ in (5), for bcc structure.

| Atom | $e_0(eV)$ | $e_1(eV)$ | $E_2(eV)$ | $\alpha$ | $\omega_e(A^3)$ |
|---|---|---|---|---|---|
| Al | -2.117 | 7.807 | 54.135 | 1.959 | 32.019 |
| Fe | -0.822 | 19.177 | 98.666 | 1.639 | 19.651 |
| V | -1.445 | 15.285 | 67.558 | 1.482 | 24.638 |
| Mo | -1.544 | 27.565 | 143.90 | 1.621 | 30.439 |

**Table 5** Pair-potential parameters for expression given by (6), found by fitting FP-LMTO total energies for the compound with the lattice under deformations.

| Atom-atom | $\phi_0(eV)$ | $r_0 (A)$ | $s_1$ | $s_2$ |
|---|---|---|---|---|
| Al-Al | -0.13113 | 2.75 | 22.36 | 6.425 |
| Fe-Fe | -0.5113 | 2.41 | 6.506 | 5.799 |
| Fe-Al | -0.4024 | 2.45 | 8.620 | 6.300 |
| V-V | -0.0492 | 2.52 | 8.821 | 6.300 |
| V-Al | -0.9630 | 2.80 | 7.656 | 4.461 |
| V-Fe | -0.1178 | 2.45 | 4.770 | 4.245 |
| Mo-Mo | -0.1537 | 2.704 | 10.163 | 8.109 |
| Mo-Al | -0.6157 | 2.800 | 8.7040 | 3.043 |
| Mo-Fe | -0.0800 | 2.450 | 5.5092 | 5.100 |

**Table 6** Three-body interaction parameters for use in the expression given by (7) found by fitting FP-LMTO total energies for distorted lattices. The cutoff distance for three-body interaction $r_c$ is equal to 3.50A.

|  | Al-Al | Fe-Fe | Fe-Al | V-V | V-Al | V-Fe | Mo-Mo | Mo-Al | Mo-Fe |
|---|---|---|---|---|---|---|---|---|---|
| $g_{ij}$ | 0.6255 | 0.4943 | 1.8306 | 1.165 | 1.831 | 0.494 | 1.451 | 1.831 | 0.494 |
| $\gamma_{ij}$ | 0.1661 | 0.0629 | 2.1530 | 1.298 | 2.153 | 0.063 | 0.6949 | 2.153 | 0.063 |